\documentclass[aps,nofootinbib,prd,showpacs,superscriptaddress,showkeys,10pt]{revtex4-1}
\usepackage[T1]{fontenc}
\usepackage[utf8]{inputenc}
\usepackage{amsmath}
\usepackage{amsfonts}
\usepackage{amssymb}
\usepackage{amsthm}
\usepackage{caption}
\usepackage{multirow}
\usepackage[english]{babel}
\usepackage{fontenc}
\usepackage{graphicx}
\usepackage{xcolor}
 \usepackage{hyperref}
\usepackage{mathptmx}
\usepackage{caption}
\usepackage{float}
\newcommand {\ignore}[1]{}

 \usepackage{comment}

\usepackage{booktabs,array}

\newcount\rowc

\makeatletter
\def\ttabular{%
\hbox\bgroup
\let\\\cr
\def\rulea{\ifnum\rowc=\@ne \hrule height 1.3pt \fi}
\def\ruleb{
\ifnum\rowc=1\hrule height 1.3pt \else
\ifnum\rowc=6\hrule height \heavyrulewidth 
   \else \hrule height \lightrulewidth\fi\fi}
\valign\bgroup
\global\rowc\@ne
\rulea
\hbox to 7em{\strut \hfill##\hfill}%
\ruleb
&&%
\global\advance\rowc\@ne
\hbox to 7em{\strut\hfill##\hfill}%
\ruleb
\cr}
\def\endttabular{%
\crcr\egroup\egroup}

\graphicspath{{./figures/}}
\begin{document}

\title{Review: Long-baseline oscillation experiments as a tool to probe High Energy Models}
\date{\today}

\author{Pedro Pasquini~$^{1,2}$}\email{pasquini@ifi.unicamp.br}

\affiliation{
$^1$ Instituto de F\'isica Gleb Wataghin - UNICAMP, {13083-859}, Campinas SP, Brazil\\
$^2$~Northwestern University, Department of Physics \& Astronomy, 2145 Sheridan Road, Evanston, IL 60208, USA}

\begin{abstract}
We review the current status of neutrino oscillation experiments, mainly focussed on T2(H)K, NO$\nu$A and DUNE. Their capability to probe high energy physics is found in the precision measurement of the CP phase and $\theta_{23}$. In general, neutrino mass models predicts correlations among the mixing angles that can be used to scan and shrink down its parameter space. We updated previous analysis and presents a list of models that contain such structure.  
\end{abstract}
\pacs{13.15.+g,14.60.St,12.60.-i,13.40.Em} 
 \maketitle

 \section{Introduction} \label{sec:prel-minim-theory}
 The upcoming sets of long-baseline neutrino experiments will stablish a new standard in the search for new physics.  Two distinct directions arrises, the  phenomenolocical approach consists on the seek of new unobserved phenomena that are present in a large class of models. They were extensively studied in the literature and are subdivided into 3 main groups: Non-Standard Interactions (NSI) searches~\cite{Guzzo:1991hi,Bolanos:2008km,Farzan:2017xzy,Ghosh:2017lim,Tang:2017qen,Liao:2016orc,Farzan:2016fmy,Blennow:2016jkn,Forero:2016ghr,Ge:2016dlx,Masud:2016gcl,Coloma:2016gei,Huitu:2016bmb}, Light Sterile Neutrinos~\cite{Boyarsky:2012rt,Heeger:2012tc,Gastaldo:2016kak,Giunti:2015mwa,Gariazzo:2015rra} and Non-unitarity~\cite{Miranda:2016wdr,Dutta:2016vcc,Dutta:2016czj,Escrihuela:2016ube,Ge:2016xya,Hernandez-Garcia:2017pwx,Das:2017fcz,C:2017scx}. The second approach is more theory based and was less explored. It focus on correlations among neutrino mixing angles predicted by high energy models. Its possibility the test of models that contains no low-energy phenomenological effects different from the Standard Model.
 
 Since the discovery of neutrino oscillations, a plethora of models was realized to tried to explain the origin of the neutrino masses. The first proposal was the See-saw mechanism~\cite{Magg:1980ut,Mohapatra:1979ia,Schechter:1980gk,Wetterich:1981bx,Foot:1988aq,Abada:2007ux} which tried to explain the smallness of neutrino masses ($m_\nu$) through a heavy mass scale ($M$) $m_\nu\propto M^{-1}$. Another possible path uses loop mechanisms, in which neutrino masses can be suppressed at zeroeth~\cite{Bonnet:2012kz} or even first order~\cite{Sierra:2014rxa}. Nevertheless, such theories usually do not explain the structure of the oscillation parameters, as they are merely free parameters. 
 
 This changes by the addition of discrete symmetry that controls the pattern of the leptonic mass matrix~\cite{King:2014nza,Haba:2001pk,Chen:2016ica}. They can predict relations among the neutrino mixing angles~\cite{CarcamoHernandez:2017kra,Dev:2017fdz,CentellesChulia:2017koy,CarcamoHernandez:2017kra,Chen:2015jta,Dicus:2010yu,M.:2014kca,Dev:2015dha,He:2015gba,Dinh:2016tuk,Ky:2016rzl,CarcamoHernandez:2017owh,Frampton:2008bz} which can be used to constrain the parameter space of such theories~\cite{Pasquini:2016kwk}.
 
 This manuscript is divided in seven section: In section~\ref{simulation} We describe current and future neutrino oscillation experiments: T2K, NO$\nu$A and DUNE and their simulation. In Section~\ref{sec:stats} we briefly discuss the statistical analysis and methods used to scan the parameter space.  In Section~\ref{sec:unconstrained} we present the sensitivity to neutrino mixing parameters expected in each experiment. In Section~\ref{sec:t23_dcp} We review the possibilities to use the $\theta_{23}-\delta_{CP}$ correlation in long-baseline experiments by updating previous analysis of two models~\cite{Chatterjee:2017xkb,Chatterjee:2017ilf}. In Section~\ref{sec:t23_t13} We review the possibility of using the $\theta_{13}-\theta_{23}$ correlation by combining long-baseline experiments with reaction measurements of $\theta_{13}$. In Section~\ref{sec:summary} we present a summary of the results.
 
\section{Long-baseline Experiments and their Simulation}\label{simulation}
 Here we choose to focus on four experimental setup, two of them are already running: T2K~\cite{Duffy:2017sfs}, NO$\nu$A~\cite{Childress:2013npa} and two had their construction approved: DUNE~\cite{Acciarri:2015uup} and T2HK~\cite{Abe:2015zbg}. Their sensitivity on the two most unknown parameters of the leptonic sector, the CP violation phase and the atmospheric mixing angle, makes them ideal to probe correlations among the mixing angles.  As it was shown in~\cite{Pasquini:2016kwk}, they can be used to shrink down the parameter space of predictive models. A short description of each experiment can be found below and on Table~\ref{tab:exp}.
\begin{table}[H]
\centering
 \begin{tabular}{ccccccc}\hline
 Experiment                       & Baseline & Size             & Target            &Expected POT                                       & Peak Energy (GeV) & Status\\ \hline \hline
 T2K~\cite{Duffy:2017sfs}         & 295 km   & 22.5 kt          & Water             & $7.8\times10^{21}$ $\left(20\times10^{21}\right)$ &  0.6              & Running (10\% total POT) \\ 
 NO$\nu$A~\cite{Childress:2013npa}& 810 km   & 14 kt            & Liq. Scintillator & $3.6\times10^{21}$                                & 2.0               & Running (17\% total POT)  \\
 DUNE~\cite{Acciarri:2015uup}     & 1300 km  & 40 kt            & Liq. Argon        & $1.47 \times 10^{21}$                             & 2.5               & Start data taking: 2026 \\
 T2HK~\cite{Abe:2015zbg}          & 295 km   & $2\times190$ kt  & Water             & $1.56 \times 10^{22}$                             & 0.6               & Start data taking: 2026 (2032)\\ \hline
 \end{tabular}
\caption{\label{tab:exp} Summary of neutrino experiments.}
\end{table}

\begin{enumerate}
\item[1.] {\underline{\bf T2K :}} The Tokai to Kamiokande (T2K) experiment~\cite{Abe:2014tzr,Duffy:2017sfs} uses the Super-Kamiokand~\cite{Abe:2016nxk} as a far detector for the J-park neutrino beam. Which consists of an off-axis (by a $2.5^\circ$ angle) predominantly muon neutrino flux with energy around 0.6 GeV. The Super-Kamiokande detector is a 22.5 kt water tank located at 295 from the J-park facility. It detects neutrino through the Cherenkov radiation emitted by a charged particle created via neutrino interaction. There is also a near detector (ND280), thus the shape of the neutrino flux is well known, and the total normalization error reaches $5\%$ for the signal and $10\%$ for the background.  T2K is already running and its current results can be found in~\cite{Haegel:2017ofz} and reaches $7\times10^{20}$ POT of flux for each neutrino/anti-neutrino mode, which corresponds to 10\% of the $7.8\times 10^{21}$ expected approved exposure. There are also plans for extending the exposure to $20\times10^{21}$ POT.

\item[2.] {\underline{\bf NO$\nu$A :}} The NuMI Off-axis $\nu_e$
  Appearance (NO$\nu$A) \cite{Patterson:2012zs,Childress:2013npa,Agarwalla:2012bv} is an off-axis (by a $0.8^\circ$ angle) that uses a neutrino beam from the Main Injector of Fermilab's beamline (NuMI). This beam consists of mostly muon neutrinos with energy around 2 GeV traveling through 810 km until arriving at the 14 kt Liquid Scintillator far detector placed at Ash River, Minnesota. The far and near detectors are highly active tracking calorimeter segmented by hundreds of PVP cells and can give a good estimate of the total signal and background within an error of $5\%$ and $10\%$ of total normalization error respectively. The planned exposure consists of a $3.6\times 10^{21}$ POT that can be achieved in 6 years of running time, working in $50\%$ in the neutrino mode and $50\%$ in the anti-neutrino mode. NO$\nu$A is already running, current results can be found in~\cite{Adamson:2017gxd,Adamson:2017qqn}.
   
\item[3.] {\underline{\bf DUNE :}}: The Deep Underground Neutrino Experiment (DUNE)~\cite{Acciarri:2016crz,Acciarri:2015uup,Strait:2016mof, Acciarri:2016ooe, Kemp:2017kbm} is a long baseline next
   generation on-axis experiment also situated in Fermilab. It flux will be generated at the LBNF neutrino beam to target a 40kt Liquid Argon time chamber projection (LarTPC) located 1300 km away from the neutrino source at Sanford Underground Research Facility (SURF). The beam consists of mostly muon neutrinos of energy around 2.5 GeV and expects a total exposure of $1.47 \times 10^{21}$ POT running 3.5 years in neutrino mode and 3.5 years in anti-neutrino mode. The Near and Far detectors are projected to obtain a total signal (background) normalization uncertainty of 4\% (10\%). The experiment is expected to start taking data around 2026.
  
\item[4.]{\underline{\bf T2HK: }} The Tokai to Hyper-Kamiokande (T2HK)~\cite{Abe:2011ts,Hyper-Kamiokande:2016dsw,Abe:2015zbg,Yokoyama:2017mnt,Migenda:2017oas} is an upgrade of the successful T2K experiment at J-Park. It uses the same beam as its predecessor T2K, an off-axis beam from the J-Park facility 295 km away from its new far detector: a two water Cherenkov tank with 190 kt of fiducial mass each. The expected total power is $1.56 \times 10^{22}$ POT to be delivered within 2.5 yrs of neutrino mode and 7.5 yrs of anti-neutrino mode in order to obtain a similar number of both neutrino types. The new design includes improvements in the detector systems and particle identification that are still in development. For simplicity, we take similar capability as the T2K experiment and will assume a 5\% (10\%) of signal (background) normalization error. The first data taking is expected to start with one tank in 2026 and the second tank in 2032. 
 \end{enumerate}
 In order to perform simulation of any neutrino experiment, the experimental collaboration uses Monte Carlo Methods, which can be performed through several event generators such as GENIE~\cite{Andreopoulos:2009rq}, FLUKA~\cite{Battistoni:2009zzb} and many others. see PDG~\cite{Patrignani:2016xqp} for a review. Such techinique requires an enormous computational power and detector knowledge, as it relies on the simulation of each individual neutrino interaction and how its products evolve inside of the detector. A simpler, but faster, simulation can be accomplished by using a semi-analitic calculation of the event rate integral~\cite{Huber:2004ka},
\begin{equation}
 N_i(\nu_\beta\rightarrow\nu_\alpha)=\int_{E_i-\Delta E_i/2}^{E_i+\Delta E_i/2}{K_{\nu_\alpha}(E,E')\phi_{\nu_\beta}(E)P_{\beta\alpha}(E)\sigma(E)dEdE'}.
\end{equation}
$N_i$ is the number of detected neutrinos with energy between $E_i-\Delta E_i/2$ and $E_i+\Delta E_i/2$. $\phi_{\nu_\beta}(E)$ describes the flux of neutrino $\nu_\beta$ arriving at the detector. $P_{\beta\alpha}$ is the oscillation probability and $\sigma(E)$ the detection cross section of the detection reaction.

$K_{\nu_\alpha}(E,E')$, also known as migration matrix, describes how the detector interpretes a $\alpha$ neutrino with energy $E$ being detected at energy $E'$ and summarizes the effect of the Monte Carlo simulation of the detector into a single function. A perfect neutrino detector is described by a delta function: $K_{\nu_\alpha}(E,E')=\delta(E-E')$ while a more realistic simulation can use a Gaussian,
\begin{equation}
 K_{\nu_\alpha}(E,E')=\frac{e^{-\frac{(E-E')^2}{2 \delta E^2}}}{\sqrt{2\pi} \delta E},
\end{equation}
where $\delta E$ parametrizes the error in the neutrino energy detection. Or a migration matrix provided by the experimental collaboration.

The public available software GLoBES~\cite{Huber:2004ka,Huber:2007ji} follows this approach and is commonly used to perform numerical simulation of neutrino experiments. There is also another tool, the NuPro packedge~\cite{NuPro} that will be publicly released soon. All the simulations in this manuscript are performed using GLoBES.


\section{Statistical Analysis and probing models: A brief Discussion}\label{sec:stats}
We are interested in a rule to distinguish between two neutrino oscillation models that can modify the spectrum of detected neutrinos in a long-baseline neutrino experiment. From the experimental point of view, one may apply a statistical analysis to quantitatively decide between two (or more) distinct hypothesis given a set of data points $H_{\rm real}$. 

Each model (${\rm M}_i$) will define a probability distribution function (p.d.f), $f\left(t| {\rm M}_i\right)$. Where the statistic test function $t$ depends on the real data points and the model parameters $\theta_i$, $i=1,2..$. The best fit of a model are defined as the values of the model parameters that maximize the p.d.f function: $f_0(M_i)={\rm Max}[f\left(H_{\rm real}, \theta^1_i| {\rm M}_i\right)]$. Thus, one can reject model $M_2$, over model $M_1$ by some certain confidence level $n$ if,
\begin{equation}
 \frac{f_0(M_2)}{f_0(M_1)}\leq C_n.
\end{equation}
$C_n$ is a constant that depend on the probability test, the number of parameters and the confidence level $n$. 

From the theoretical point of view, the real data points were not yet measured, this means that in order to find the expected experimental sensitivity we need to produce pseudo-data points $H_{\rm real}$ by adding an extra assumption on which model is generating the yet-to-be-measured data points. That means there are various ways of obtaining sensitivity curves, each of them are described in Table~\ref{tab:hypotesis}.
\begin{table}[H]
 \centering
 \begin{tabular}{cccc}\hline
  Cases   & pseudo-data & Null Hypothesis & Test Hypotesis\\ \hline \hline
 General  &     M$_i$   & M$_1$ & M$_2$   \\
    I     &     Standard-3$\nu$ & Standard-3$\nu$ & New Model   \\
   II     &     New Model       & Standard-3$\nu$ & New Model   \\
  III     &     Standard-3$\nu$ & New Model       & Standard-3$\nu$  \\
   IV     &     New Model       & New Model       & Standard-3$\nu$ \\ \hline
 \end{tabular}
 \caption{\label{tab:hypotesis} Description of the possible hypotesis taken to generate the numerical anallysis. Here, Standard-3$\nu$ means the standard 3 neutrino oscillation.}
\end{table}
Although one can always generate the pseudo-data points using any desired model at any point in its parameter space, the usual approach is to assume that the data points are generated by the standard 3 neutrino oscillation (Standard-3$\nu$) model with parameters given by current best fit values. We will use this approach in the work. Current best fit values are described in Table~\ref{tab:3nubest} and were taken from~ \cite{deSalas:2017kay}.
\begin{table}[H]
 \centering
 \begin{tabular}{cccc}\hline
  parameter  & value & error\\ \hline \hline
    $\Delta m_{21}^2/10^{-5}$ & 7.56 $\rm eV^2$ & (19)  \\ 
    $\Delta m_{31}^2/10^{-3}$ & 2.55 $\rm eV^2$ & (4) \\ 
    $\sin^2\theta_{12}$       & 0.321           & (18) \\ 
    $\sin^2\theta_{13}$       & 0.02155         & (90)\\ 
    $\sin^2\theta_{23}$       & 0.430           & (20) \\     
    $\delta_{\rm CP}/\pi$     & 1.40            & (31) \\ \hline
 \end{tabular}
 \caption{\label{tab:3nubest} Current best fit values of Standard-3$\nu$ as given by~ \cite{deSalas:2017kay}. Notice that Normal Hierarchy is assumed.}
\end{table}

\subsection{Frequentist Analysis}
The chi-square test~\cite{James:2006zz,William:1952,Patrignani:2016xqp} is the most commom statistical analysis choosen to teste the compatibility between the experimental data and the expected outcome of a given neutrino experiment. It bases on the construction of a Gaussian chi-squared estimator ($\chi^2$) so that $f(t|Model)=Ne^{-\chi^2/2}$. This means that the best fit values are obtained by the set of values that globaly minimizes the function $\chi^2$. For long-baseline neutrino oscillation experiments the chi-square function can be devided into three factors,
\begin{equation}
 \chi^2=\chi^2_{\rm data}+\chi^2_{\rm sys}+\chi^2_{\rm prior}.
\end{equation}
Where $\chi^2_{\rm data}$ in the simplest case reduces to a Poissonian Pearson's statistic
\begin{equation}\label{eq:simplechi}
\chi^2_{\rm data}=\sum_i^N \left[\frac{N_i^{\rm obs}-(1-a)N_i^{s}-(1-b)N_i^{b}}{\sqrt{N_i^{\rm obs}}}\right]^2.
\end{equation}
$N_i^{\rm obs}$ is the number of observed neutrino in the bin $i=1,2,3..N$ and are the pseudo-data points generated by a given model. $N_i^{s}$ ($N_i^{b}$) is the signal (background) observed neutrinos as expected by a given model and depend on the model parameters. The $\chi^2_{\rm sys}$ comprises the experimental uncertanties and systematics. For the $\chi^2$ in Eq.~\ref{eq:simplechi}, it is given by,
\begin{equation}
 \chi^2_{\rm sys}=\left(\frac{a}{\sigma_a}\right)^2+\left(\frac{b}{\sigma_b}\right)^2.
\end{equation}
Here, $\sigma_a$ ($\sigma_b$) is the total normalization error in the signal (background) flux. Finally, $\chi^2_{\rm prior}$ contains all the prior information one wishes to include in the model parameters. In this work we will assume $\chi^2_{\rm prior}=0$ unless stated otherwise.

The exponential nature of the chi-squared estimator makes it straigntfoward to find the confidence levels for the model parameters. It sufices to define the function,
\begin{equation}
 \Delta\chi^2=\chi^2_{\rm min}(\theta_i|M_2)-\chi^2_{{\rm min}}(M_1),
\end{equation}
where $\chi^2_{{\rm min}}(M_1)$ is the chi-squared function assuming model $M_1$ calculated in its best fit and $\chi^2_{\rm min}(\theta_i|M_2)$ is the chi-squared function assuming model $M_2$ minimized over all the desired free parameters. Thus, the confidence levels are obtained by finding the solutions of
\begin{equation}
 \Delta\chi^2\leq A_n.
\end{equation}
$\theta_i$ are all the fixed parameters of model $M_2$ and $A_n$ are the constants that define the probability cuts and depend on the number of parameters in $\chi^2(\theta_i|M_2)$ and the confidence probability. For $n\sigma$ intervals and one parameter, $A_n=n^2$.

Notice that $\Delta\chi^2$ is in fact a function of the parameters one assumes to generate the pseudo-data points, which we call {\it True Values} and denote $\theta_i (\rm True)$ and the parameters of the model we wish to test, which we call {\it Test Values} and denote $\theta_i(\rm Test)$.
\section{Measurement of Oscillation parameters in Long-baseline Experiments}\label{sec:unconstrained}
 The main goal of long-baseline experiments is to measure with high precision the two most unknown oscillation parameters: the CP phase and the atmospheric mixing angle $\theta_{23}$ through the measurement of the neutrino/anti-neutrino $\nu_\mu\rightarrow\nu_\mu$ survival and $\nu_\mu\rightarrow\nu_e$ transition of neutrinos from the beamline. In special, only the transition is sensitive to $\delta_{\rm CP}$ and described, to first order in matter effects, by the probability function below.
 \begin{align}
  P\left(\nu_\mu\rightarrow\nu_e\right)=&4c_{13}^2s^2_{13}s_{23}^2 \sin^2\Delta_{31}\\
  &+8c_{13}^2s_{12}s_{13}s_{23}(c_{12}c_{23}\cos\delta_{\rm CP}-s_{12}s_{13}s_{23})\cos\Delta_{32}\cos\Delta_{31}\cos\Delta_{21}\\
  &-8c_{13}^2c_{12}c_{23}s_{12}s_{13}s_{23}\sin\delta_{\rm CP}\sin\Delta_{32}\sin\Delta_{31}\sin\Delta_{21}\\
  &+4s_{12}^2c_{13}^2(c_{12}^2c_{23}^2+s^2_{12}s_{23}^2s_{13}^2-2c_{12}c_{23}s_{12}s_{23}s_{13}\cos\delta_{\rm CP})\sin^2\Delta_{21}\\
  &-8c_{13}^2s_{13}^2s_{23}^2\frac{aL}{4E}(1-2\sin_{13}^2)\cos\Delta_{32}\sin\Delta_{31}\\
  &+8c_{13}^2s_{13}^2s_{23}^2\frac{a}{\Delta m^2_{31}}(1-2\sin_{13}^2)\sin^2\Delta_{31}.
 \end{align}
Where $c_{ij}=\cos\theta_{ij}$, $s_{ij}=\sin\theta_{ij}$, $\Delta_{ij}=\Delta m_{ij}^2L/4E$ and $a=2\sqrt{2}G_{\rm F}n_eE$. $G_{\rm F}$ is the fermi constant and $n_e$ is the electron density in the medium. $E$ is the neutrino energy and $L$ the baseline of the experiment and are choosen to obey $L/E\sim 500$ in order to enhance the effect of the CP phase. The anti-neutrino probability is obtained by change $a\rightarrow-a$ and $\delta_{\rm CP}\rightarrow-\delta_{\rm CP}$. Thus, the difference betweem neutrino and anti-neutrino comes from matter effects and the CP phase. 
 \begin{figure}[!ht]
\centering
 \includegraphics[scale=.38]{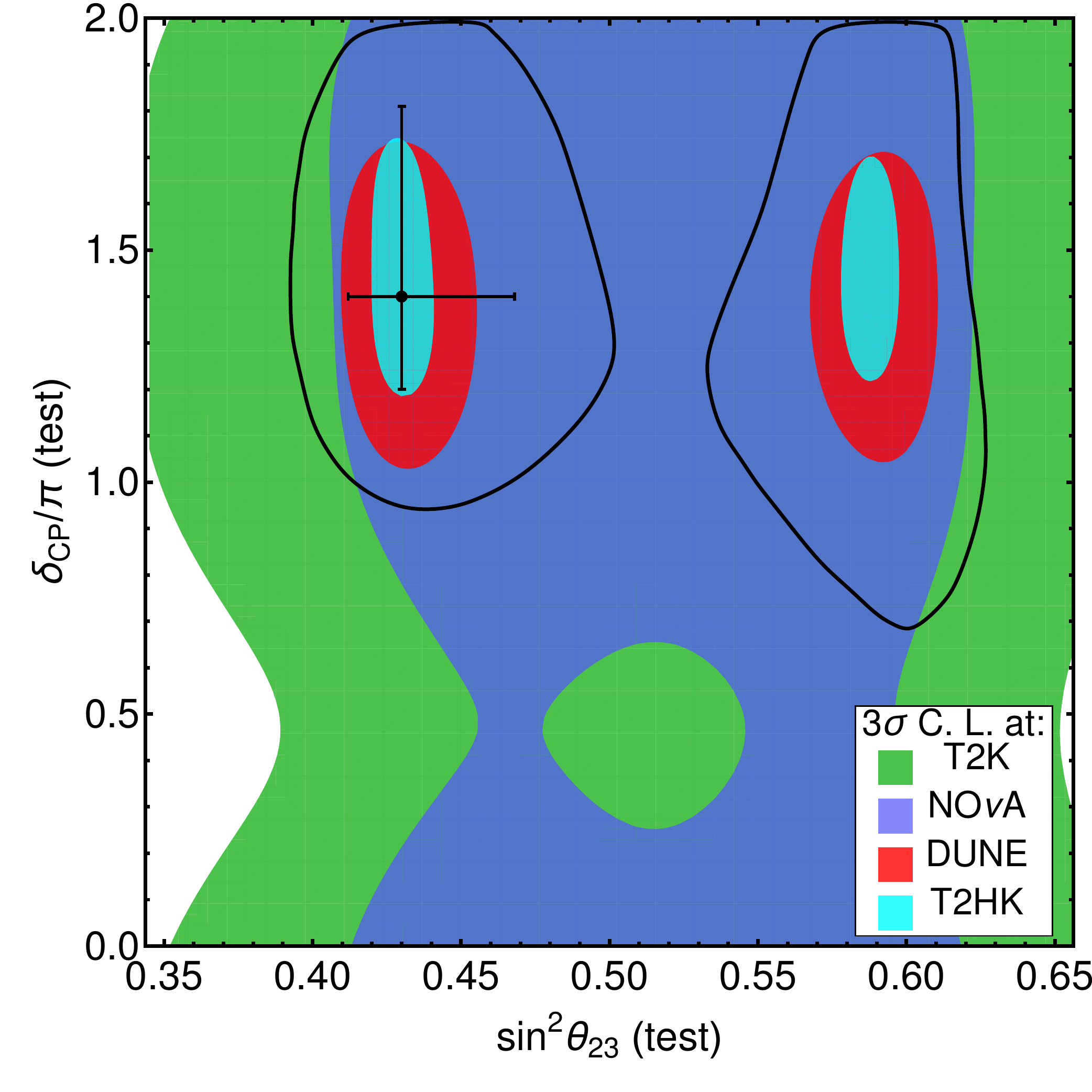}
\caption{Expected sensitivity regions of $\sin^2\theta_{23}(test)$ versus  $\delta_{CP}(test)$ assuming as the true and test model the standard-3$\nu$ paradigm for the three long-baseline detectors discussed: (1) T2K (Green), (2) NO$\nu$A (Blue), (3) DUNE (Red) and (4) T2HK (Cyan). The black curve is current 90\% C. L. and the black point is the current best fit given in Table~\ref{tab:3nubest}. Notice that within this assumptions the octant would remain unresolved even at 3$\sigma$ C.L.}
  \label{precision}
  \end{figure}
  It turns out that the T2HK is the most sensitivity to $\delta_{CP}$ as it has a bigger statistic and lower matter effect, and can reach a $8\sigma$ difference between CP conservation and maximal CP-non-conservation~\cite{Hyper-Kamiokande:2016dsw}, in contrast with DUNE's $5.5\sigma$~\cite{Acciarri:2016crz}. In Fig.~\ref{precision} we plotted the expected allowed regions of $\theta_{23}({test})$ versus $\delta_{\rm CP}({test})$ at $3\sigma$ for each experiment. We assumed the true value of the parameters as those given in Table~\ref{tab:3nubest}. The black region is the current 90\% C. L. region and the black points are the best-fit point. T2HK is the most sensitive experiment in reconstructing both parameters, followed by DUNE. NO$\nu$A and T2K are the first experiments to measure a difference between matter and anti-matter in the leptonic sector, but cannot measure the CP phase with more than 3$\sigma$.
 \begin{figure}[!ht]
 \centering
  \includegraphics[scale=.42]{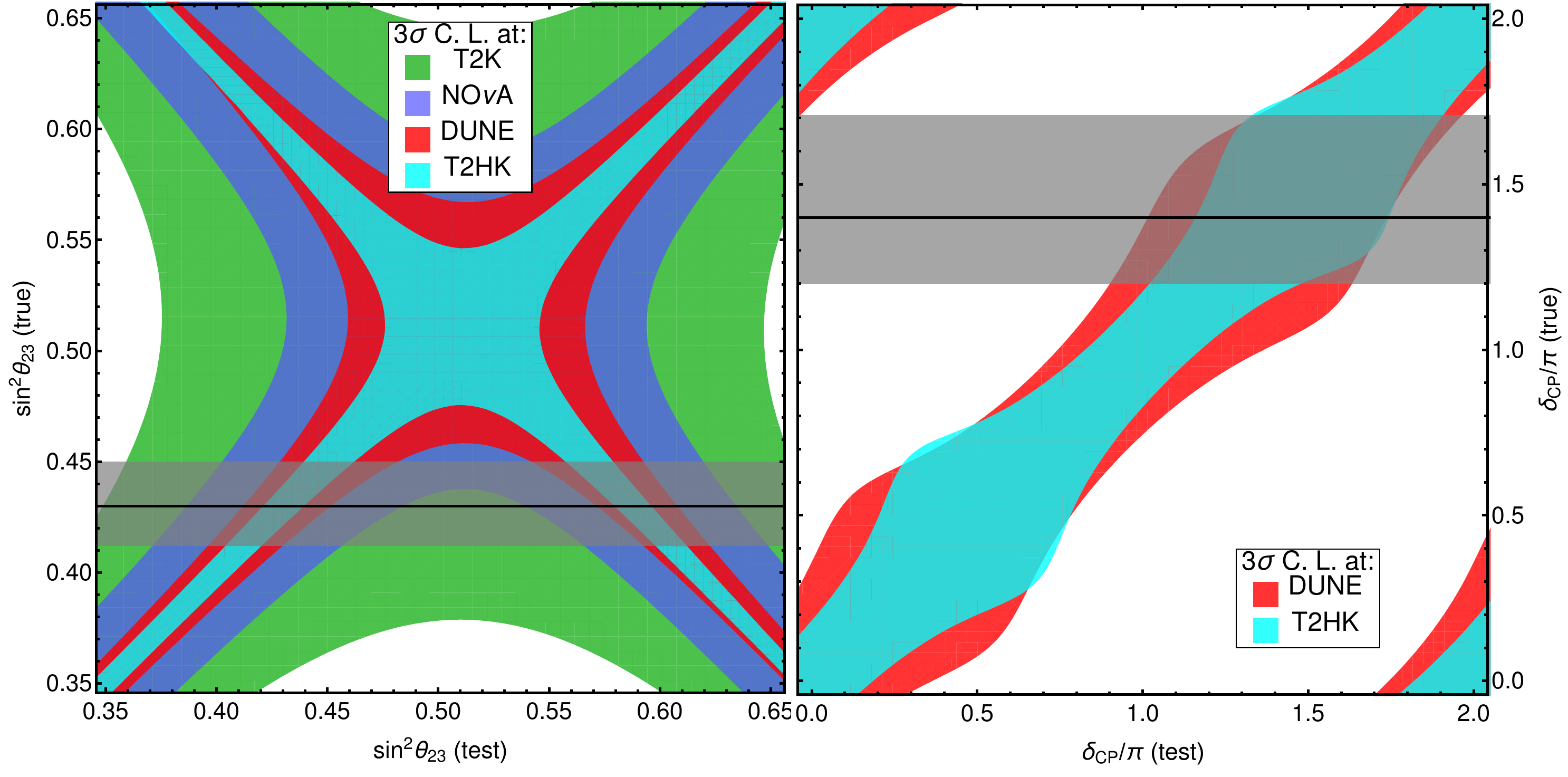}
  \caption{The left (right) panel corresponds to the expected reconstruction of
    the oscillation parameter $\theta_{23}$ ($\delta_{CP}$). The black line
    indicates the best fit value given in Table~\ref{tab:3nubest} and the gray area corresponds to its $1\sigma$ region. The colored areas represents the regions that the experiments cannot distinguish within more than 3$\sigma$ for:  (1) T2K (Green), (2) NO$\nu$A (Blue), (3) DUNE (Red) and (4) T2HK (Cyan). In the right panel we did not included T2K or NO$\nu$A as they cannot reconstruct the CP phase with more than 3$\sigma$.}
 \label{octant_sensitivity2}
  \end{figure}
Notice that the experiments cannot discover the correct octant of $\theta_{23}$ at $3\sigma$, that is, they cannot tell if $\theta_{23}>\pi/4$ (High Octant) or $\theta_{23}<\pi/4$ (Lower Octant) unless they are supplemented by an external prior. This effect is independent of the value of $\theta_{23}$ as can be observed in Fig.~\ref{octant_sensitivity2}-Left panel where we plotted the reconstruction of the $\theta_{23}(test)$ given a fixed true value of $\theta_{23}(true)$ of each experiment. The black line corresponds to current best-fit and the gray area is the 1$\sigma$ region. The x-like pattern of the region shows that given any true value of $\theta_{23}$ there is $3\sigma$ region in the correct octant and in the wrong octant. Nevertheless, the octant can be obtained if one incorporates a prior to the $\theta_{13}$ angle~\cite{Nath:2015kjg,Bora:2014zwa,Minakata:2002jv} and future prospects on the measurement of $\theta_{13}$ by reactor experiments will allow both DUNE and T2HK to measure the octant if the atmospheric angle does not all inside the region $0.47<\sin^2\theta_{23}<0.53$~\cite{Chatterjee:2017irl}. 

For completeness, we show in the left panel of Fig.~\ref{octant_sensitivity2} the reconstruction of the $\delta_{\rm CP}(test)$ given a fixed true value of $\delta_{CP}(true)$. The black line represents current best fit and the gray area the $1\sigma$ region. We don't show the plots for NO$\nu$A or T2K as they cannot reconstruct the CP phase at $3\sigma$. The sensitivity is a little bit worse around maximum CP violation $\delta_{\rm CP}=\pi/2$ or $3\pi/2$ but in general it does not change much when one varies the $\delta_{\rm CP}(true)$.
 
\section{\texorpdfstring{$\theta_{23}$}{t23} and \texorpdfstring{$\delta_{CP}$}{dcp} Correlation and Probing Models}\label{sec:t23_dcp}

In spite of being relatively low energy (< few GeV), neutrino experiments can be a tool to probe high energy physics. Many neutrino mass models predicts relations such as neutrino mass sum rules~\cite{Spinrath:2016ldz,Buccella:2017jkx,Gehrlein:2016wlc} that can be probed in neutrinoless double beta decay~\cite{King:2013psa} and relations among the neutrino mixing parameters. To name a few examples we cite~\cite{CarcamoHernandez:2017kra,Dev:2017fdz,CentellesChulia:2017koy,CarcamoHernandez:2017kra}. They can be put to test by a scan of the parameter space much like it was done by the LHC in search for new physics. Thus, inspired by the precision power of future long-baseline neutrino experiments, it was shown in~\cite{Pasquini:2016kwk} that models that predict a sharp correlation between the atmospheric angle and the CP phase can be used to put stringent bounds on parameters of such models. 

\begin{figure}[!ht]
    \centering
\includegraphics[scale=0.41]{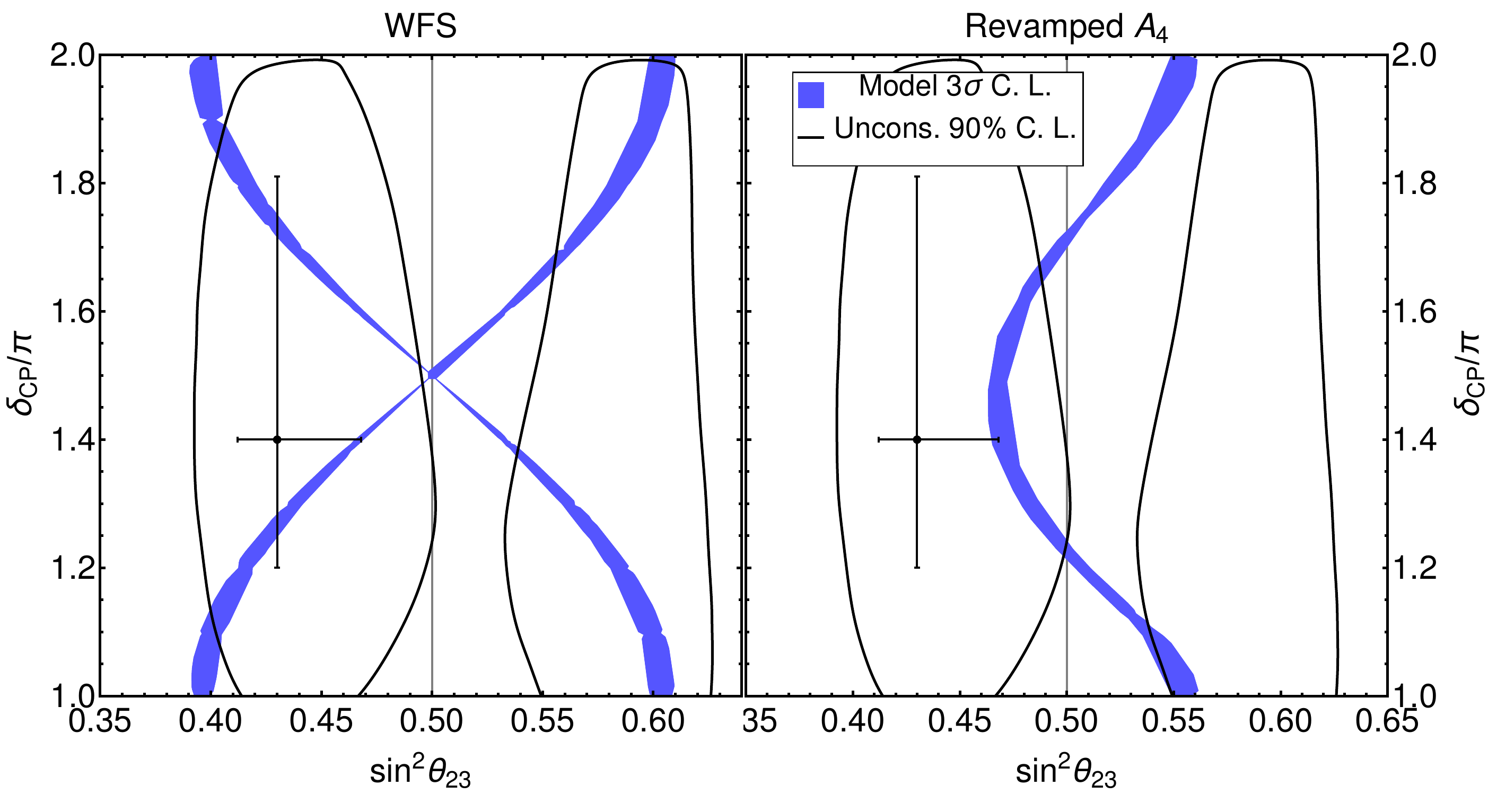}
\caption{\label{fig:current1} 
In Blue: Possible parameter values allowed by the two benchmark models: Warped Flavour Symmetry (Left) and Revamped $A_4$-BMV (Right). The regions are constructed by varying all the free paramters of the model and selecting those that are allowed at $3\sigma$ in current global fit analysis~\cite{deSalas:2017kay}. The black line corresponds to curren 90\% C. L. region and the black-dot is the best-fit of Table~\ref{tab:3nubest}. Normal Hierarchy is assumed.}
 \end{figure}

In general, a predictive neutrino mass model ${\cal M}$ is constructed by imposing a symmetry on the Lagrangian and can be parametrized by a set of free parameters $\phi_i$, $i=1,2,...N$ which can be translated into the usual neutrino mixing parameters from the neutrino mass matrix, that is,
\begin{align}
 \theta_{jk}\equiv&\theta_{jk}(\phi_i),\\
 \delta_{\rm CP}\equiv&\delta_{\rm CP}(\phi_i).
\end{align}
Because of the symmetry on the Lagrangian, not all possible mass matrices are allowed to be generated and the free parameters $\phi_i$ may not span the entire space of the mixing parameters $\theta_{ij}$ and $\delta_{\rm CP}$. Thus, in principle, it is possible to probe or even exclude a model if the real best fit falls into a region that the model ${\cal M}$ cannot predict. As an example, in Fig.~\ref{fig:current1} we plot the allowed parameter space o two discrete symmetry based models, the Warped Flavour symmetry (WFS) model~\cite{Chen:2015jta} - Left and the Revamped $A_4$ Babu-Ma-Valle (BMV) model~\cite{Morisi:2013qna} - Right. The black curves represents currently unconstrained (Standard-$3\nu$) 90\% C. L. regions for the neutrino parameters and the black point the best-fit value, while the blue region represents the $3\sigma$ allowed parameter space of the two models.
 Notice that even for the 3$\sigma$ range the model can only accommodate a much smaller region than the unconstrained. This is a reflex of the symmetries forced upon those models by construction, in WFS a maximal CP phase implies a $\theta_{23}=\pi/4$ and the smaller the CP violation, the farther away from $\pi/4$ the atmospheric angle is. While in BMV a maximal CP phase implies a lower octant atmospheric mixing and it can't fit a $|\theta_{23}-\pi/4|>0.02\pi$.
 
 By using this approach, a full scan of the parameter space was performed for those two models, in~\cite{Chatterjee:2017xkb} for the WFS model and in~\cite{Chatterjee:2017ilf} for the Revamped $A_4$ model. 
 
 \begin{figure}[!ht]
 \centering
 \includegraphics[scale=.3]{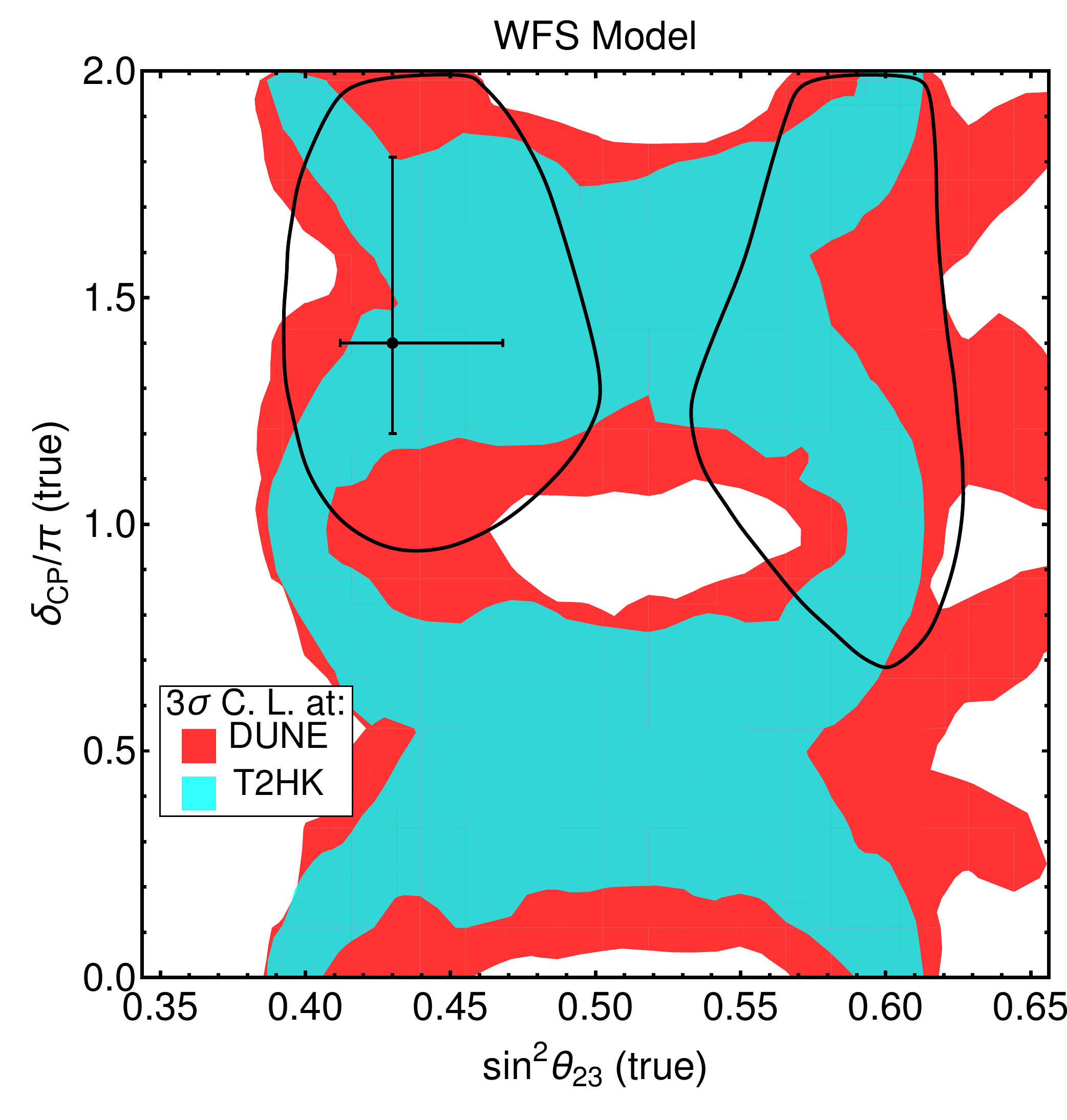}
\includegraphics[scale=0.3]{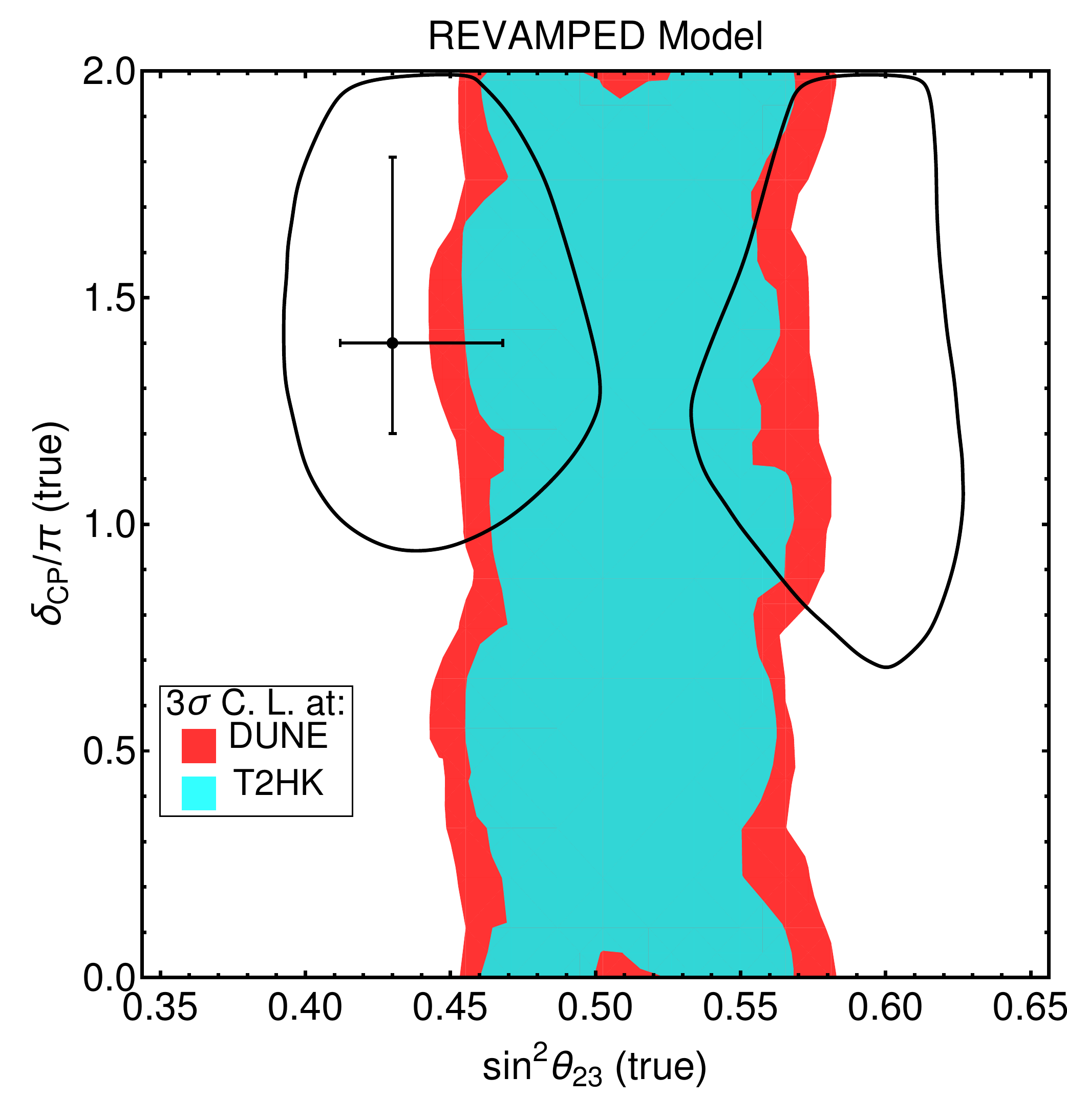}
\caption{\label{fig:scan} 
Expected sensitivity regions at which DUNE (Red) or T2HK (Cyan) would not exclude the WFS model (Left) and the revamped BMV model (Right) at $3\sigma$ confidence level. The black contours correspond to 90\% C.L. of current global-fit~\cite{deSalas:2017kay}.  
  }
   \end{figure}
We show in Fig.~\ref{fig:scan} an updated version of their results. The colored regions represent regions of the parameter space in which the model {\bf cannot} be excluded with more than $3\sigma$ for DUNE (red) and T2HK (blue) experiment, both T2K and NO$\nu$A cannot probe the CP phase with more than 3$\sigma$, thus, they cannot exclude the model alone. 

This means that if future long-baseline experiments measure a specific combination of $\delta_{\rm CP}$ and $\theta_{23}$ as its best fit that does not fall into the colored regions, they may be able to exclude the model. Therefore, those kind of analysis are guidelines to decide which model can or cannot be tested given the future results of DUNE and T2HK and are worth to be performed in any model that contains predictive correlations among the CP phase and the atmospheric mixing, such as~\cite{CarcamoHernandez:2017kra,Dev:2017fdz,CentellesChulia:2017koy,CarcamoHernandez:2017kra,Srivastava:2017sno} and many others. It is also worth mention that combination of long-baseline measurements and reactors can greatly improve the sensitivity of the analysis.
\section{\texorpdfstring{$\theta_{13}$}{t13} and The atmospheric Octant}\label{sec:t23_t13}
The analysis in the last section can be extended to include another type of correlation that tries to explain the smallness of the reactor angle $\theta_{13}\sim O(10^\circ)$. A general approach common in many models~\cite{Dicus:2010yu,M.:2014kca,Dev:2015dha,He:2015gba,Dinh:2016tuk,Ky:2016rzl,CarcamoHernandez:2017owh,Frampton:2008bz} imposes a given symmetry on the mass matrix that predicts $\theta_{13}=0$. Which is later spontaneously broken to give a small correction $\delta\theta_{13}\sim O(10^\circ)$ to the reactor angle. It turns out that in order to generate non-zero $\theta_{13}$ one automatically generates corrections to other mixing angles $\delta\theta_{ij}\neq0$.

This can be easely observed by considering a toy model that predicts the Tri-bi-maximal mixing matrix,
\begin{equation}
 U_{PMNS}=U_{TBM}=\left(
 \begin{array}{ccc}
  \sqrt{\frac{2}{3}} &  \sqrt{\frac{1}{3}} & 0\\
    \sqrt{\frac{1}{3}} &   -\sqrt{\frac{1}{3}} &   \sqrt{\frac{1}{2}}\\
        -\sqrt{\frac{1}{3}} &   \sqrt{\frac{1}{3}} &   \sqrt{\frac{1}{2}}
  \end{array}
 \right).
\end{equation}
Any consistent small correction to the mixing matrix should maintain its unitary character. In special, we can set a correction in the $2-3$ plane via the matrix
\begin{equation}
 \delta U_{23}=\left(
 \begin{array}{ccc}
  1&  0 & 0\\
 0 &   1 &   \delta\theta\\
 0 &  -\delta\theta &  1
  \end{array}
 \right).
\end{equation}
Notice that $ \delta U_{23}. \delta U_{23}^\dagger=1+O(\delta\theta^2)$. If we change the mixing matrix\footnote{Notice that the correction $\delta U_{23}.U_{TBM}$ cannot produce a non-zero $\theta_{13}$} by $U_{PMNS}=U_{TBM}\rightarrow U_{TBM}.\delta U_{23}$ then $\theta_{13}=\frac{1}{\sqrt{3}}\left|\frac{\pi}{4}-\theta_{23}\right|$. The general case can be described by a initial mixing matrix $U_{PMNS}=U_0$ that is later corrected by a rotation matrix $U_{ij}$,
\begin{equation}
 U_{PMNS}=U_0\rightarrow U_{PMNS}=U_{ij}.U_0 \quad {\rm or}\quad U_0.U_{ij}
\end{equation}
All the possible combinations of corrections from Tri-Bi-MAximal, Bi-Maximal and democratic mixing were considered in~\cite{Chao:2011sp}. In special, one can investigate a general correlation of $\theta_{13}$ to the non-maximality of the atmospheric angle,
\begin{equation}\label{Eq:t13corre}
 \theta_{13}=F\left(\delta \theta_{23}\right).
\end{equation}
Where $F$ is a function of the correction $\delta \theta_{23}=\left|\frac{\pi}{4}-\theta_{23}\right|$. Long-baseline experiments alone are not too sensitive to changes in the reactor angle, nevertheless, it was shown in~\cite{Pasquini:2017hby} that it is possible to use such correlation to probe the parameter space of such models by combining long-baseline and reactor experiments.

This can be accomplished in a model-independend approach by series expanding Eq.~\ref{Eq:t13corre},
\begin{equation}\label{eq:t13_corre_eq}
 \theta_{13}=f(0)+f'(0)\left|\frac{\pi}{4}-\theta_{23}\right|\equiv \theta^0_{13}+f\left|\frac{\pi}{4}-\theta_{23}\right|.
\end{equation}
This encompass both the uncorrelated (standard-3$\nu$) case if one sets $f=0$ and assumes $\theta^0_{13}$ as a free parameter and the small correction case by setting $\theta^0_{13}=0$ and $f\neq0$. On table~\ref{tab:pred} we present many models that contain this kind of correlation and their possible parameters values for $f$ and $\theta^0_{13}$.
\begin{table}[H]
\centering
\begin{tabular}{ccc}
 \hline
 Model & $f$  & $\theta^0_{13} [{\rm rad}]$ \\ \hline \hline
 \cite{Frampton:2008bz} & $\sqrt{2}$ & 0\\
  \cite{Ky:2016rzl,Dinh:2016tuk}& $0.35$ & [0,0.35] \\
 \cite{He:2015gba} & $0.1$ or $10$ & 0.62\\
 \cite{CarcamoHernandez:2017owh} & $1/\theta_0$ & [-1,1]  \\ 
 $U_{13}.U_{TBM}$ & $6.3$ & $0$  \\
 $U_{12}.U_{TBM}$  & $6.3$ & $0$ \\
  $U_{TBM}.U_{23}$  & $1/\sqrt{3}$ & $0$ \\\hline
\end{tabular}
\hspace{1cm}
\begin{tabular}{ccc}
 \hline
 Model & $f$  & $\theta^0_{13} [{\rm rad}]$ \\ \hline \hline
 $U_{BM}.U_{23}U_{13}$ & $1/2$& $0$  \\
 $U_{TBM}.U_{23}U_{12}$& $2$ & $0.157$  \\
$U_{TBM}.U_{23}U_{13}$ & $1/\sqrt{2}$ & $0$  \\
$U_{TBM}.U_{13}U_{12}$ & $2/\sqrt{2}$ & $0$  \\
$U_{BM}.U_{13}U_{12}$ & $\sqrt{3/2}$  & $0$ \\
$U_{BM}.U_{23}U_{12}$ & $\sqrt{3/2}$  & $0$ \\
$U_{BM}.U_{23}U_{13}$ & $1/2$& $0$  \\ \hline
\end{tabular}
\caption{\label{tab:pred} Summary of models containing reactor and atmospheric angle correlation. All the possible combinations of corrections from Tri-Bi-MAximal, Bi-Maximal and democratic mixing were considered in~\cite{Chao:2011sp}.} 
\end{table}

\begin{figure}[!ht]
\centering
\includegraphics[scale=.4]{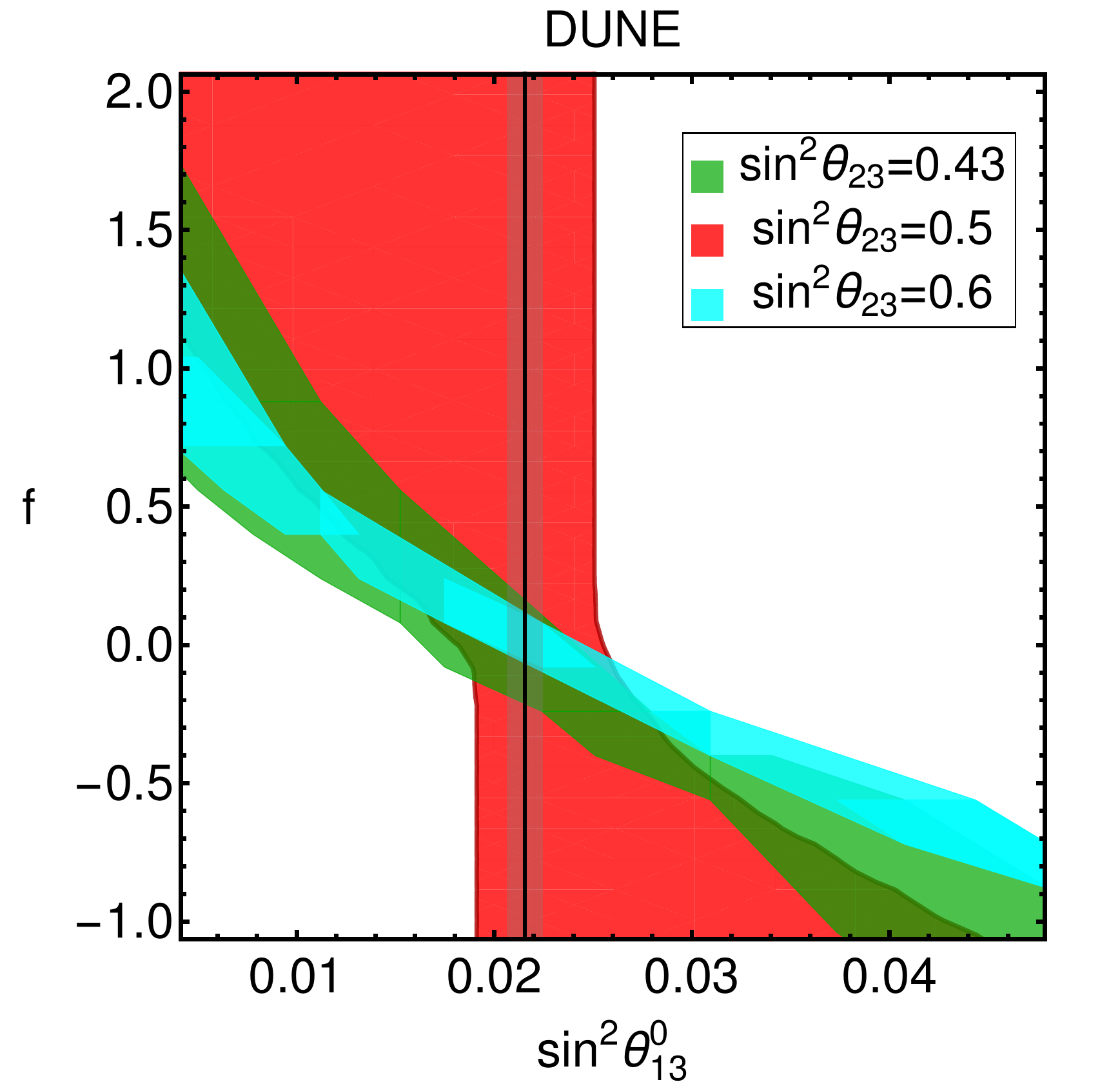}
\includegraphics[scale=.4]{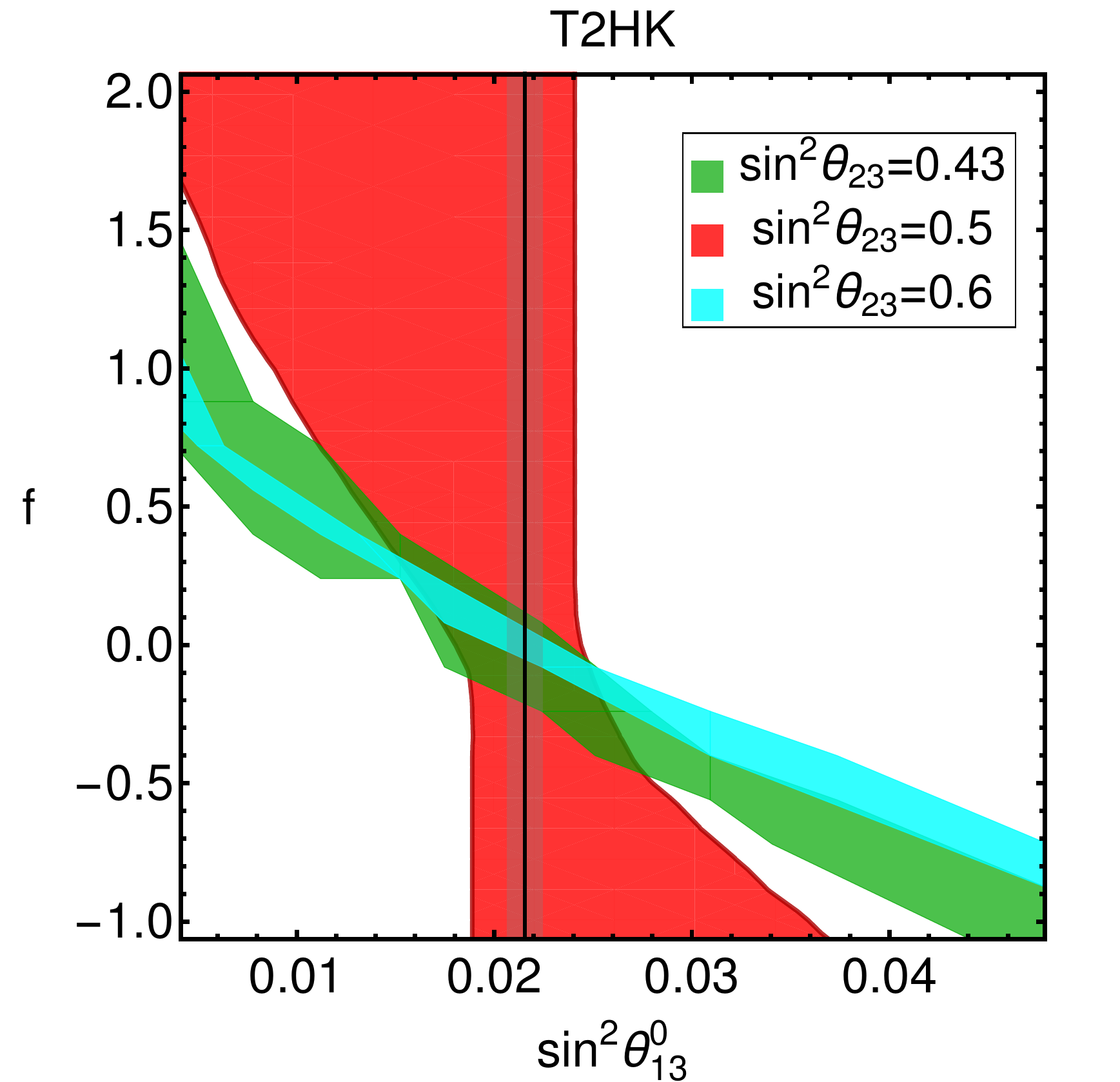}
\caption{\label{fig:plot_general2} General parameter space regions that cannot be distinguished from the unconstrained relation hypothesis at more than $3\sigma$ by future long-baseline neutrino oscillation experiments in combination with reactor measurements as a function of the two parameters of Eq.~\ref{eq:t13_corre_eq}: $\theta_{13}^0$ and $f$. The analysis assumed the central value of the reactor angle as the best-fit given in Table~\ref{tab:3nubest} and three values for the atmospheric angle: (1) $\sin^2\theta_{23}(true)=0.43$ (Green), (2) $\sin^2\theta_{23}(true)=0.5$ (Red) and (3) $\sin^2\theta_{23}(true)=0.6$ (Cyan) for DUNE (Left) and T2HK (Right). }
\end{figure}

On Fig.~\ref{fig:plot_general2} we update the potential exclusion regions where models of the form $\theta_{13}^0=0$ can be excluded for each value of $\sin^2\theta_{23}(true)$ at 3$\sigma$ by DUNE+reactors and T2HK+reactors. The true value of $\theta_{13}$ is set to the central value of Table~\ref{tab:3nubest} and its error is assumed to be $3\%$. The colored regions represent the regions that {\bf cannot} be excluded with more than 3$\sigma$. There we can see that models that contain strong correlations ($f>1.9$) or weak correlations ($f<0.8$) can be excluded from any set of atmospheric angle.

\begin{figure}[!ht]
\centering
\includegraphics[scale=.4]{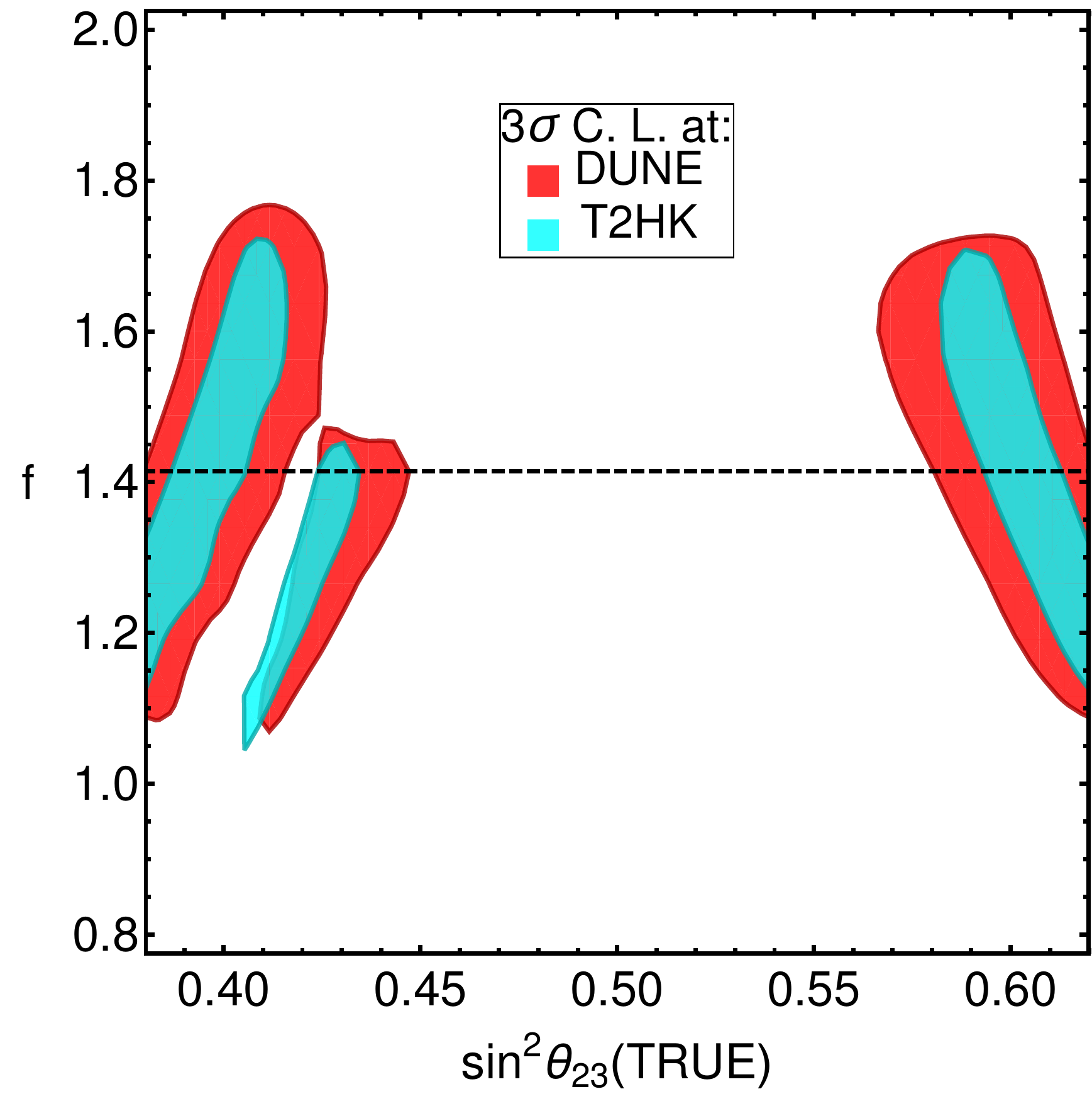}
\caption{\label{fig:plot_general} Regions that future neutrino long-baseline oscillation experiments cannot exclude the models that follow Eq.~\ref{eq:t13_corre_eq} at more than 3$\sigma$ C.L. as a function of the true value of the atmospheric mixing angle for: DUNE(Red) and T2HK(Cyan)}
\end{figure}

The general case for any $\theta_{13}^0$ is presented in Fig.~\ref{fig:plot_general2} for DUNE on the left panel and for T2HK on right panel for three values of $\sin^2\theta_{23}$: 0.43 (Green), 0.5 (Blue) and 0.6 (Red). The region shrinks greatly as the true value of the atmospheric angle goes away from the maximal mixing $\theta_{23}(true)=\pi/4$.

\section{Summary}\label{sec:summary}
The state-of-art of long-baseline neutrino oscillation experiments are T2(H)K, NO$\nu$ and DUNE. They will be capable of reaching very good precision in the reactor and atmospheric mixing angle and will measure for the first time the CP violation phase. This will create an opportunity to put at test a plethora of neutrino mass models that predict values and correlations among the parameters of the PMNS matrix~\cite{Pasquini:2016kwk,Chatterjee:2017xkb,Chatterjee:2017ilf,Srivastava:2017sno,Ge:2011ih,Ge:2011qn}.

Here we briefly discuss the fitting approach that quantify the ability of long-baseline experiments to exclude predictive high energy models. Two types of correlations can be used: The $\theta_{23}-\delta_{\rm CP}$ correlation is found in many models containing a variety of symmetries~\cite{CarcamoHernandez:2017kra,Dev:2017fdz,CentellesChulia:2017koy,CarcamoHernandez:2017kra,Chen:2015jta}. Nevertheless, 
each model in the market may contain a different correlation, and most models are still in need to be analyzed. On the other hand, the $\theta_{13}-\theta_{23}$ correlation can only be probed by combining Long-baseline with reactor experiments, as the former are not sensible enough to $\theta_{13}$ variations. However, we can take a model-independent approach~\cite{Pasquini:2017hby} that covers most models that try to explain the smallness of the $\theta_{13}$ angle trough an spontaneous symmetry breaking~\cite{Dicus:2010yu,M.:2014kca,Dev:2015dha,He:2015gba,Dinh:2016tuk,Ky:2016rzl,CarcamoHernandez:2017owh,Frampton:2008bz}. We present a set of figures~\ref{fig:scan}, \ref{fig:plot_general} and \ref{fig:plot_general2} containing the potential exclusion regions of each model here analysed that can be used as a benchmark when the future experiments starts to run.
\section*{Acknowledgments}

P. P. was supported by FAPESP grants 2014/05133-1, 2015/16809-9, 2014/19164-6 and FAEPEX grant N. 2391/17. Also, through the APS-SBF collaboration scholarship.

 \bibliographystyle{apsrev4-1}

\end{document}